\documentclass[%
 reprint,
longbibliography,
superscriptaddress,
 amsmath,amssymb,
 aps,
prb,
]{revtex4-2}

\usepackage{graphicx}
\usepackage{dcolumn}
\usepackage{bm}

\usepackage{float}
\usepackage{bbold}
\usepackage{latexsym}    
\usepackage{wasysym}
\usepackage{amsfonts}
\usepackage{siunitx}
\usepackage{blindtext}
\usepackage{caption} 
\renewcommand\Re[1]{\operatorname{Re} \left( #1 \right)}
\renewcommand\Im[1]{\operatorname{Im} \left( #1 \right)}

\newcommand{\dg}[1]{\ensuremath{#1^\circ}}

\newcommand{\nmetr}[1]{\ensuremath{#1~\text{nm}}}

\newcommand{\kvec}{\ensuremath{\vec{k}}}

\newcommand{\epsi}{\ensuremath{\varepsilon}}

\newcommand\T{\ensuremath{\mathcal{T}}}
\newcommand\PP{\ensuremath{\mathcal{P}}}
\newcommand\R{\ensuremath{\mathcal{R}}}
\newcommand\A{\ensuremath{\mathcal{A}}}
\newcommand\E{\ensuremath{\mathcal{E}}}
\newcommand\HH{\ensuremath{\mathcal{H}}}

\newcommand\SSS{\ensuremath{\mathcal{S}}}

\newcommand{\MoO}{\ensuremath{\text{MoO}_{\text{3}}}}

\newcommand{\SiO}{\ensuremath{\text{SiO}_{\text{2}}}}

\newcommand{\wLO}{\ensuremath{\omega_{\text{LO}}}}

\newcommand{\rpp}{\ensuremath{r_{pp}}}

\begin{document}


\title{Layer-Resolved Resonance Intensity of Evanescent Polariton Modes in Anisotropic Multilayers}

\author{Nikolai Christian Passler}
\affiliation{Fritz-Haber-Institut der Max-Planck-Gesellschaft, Faradayweg 4-6,14195 Berlin, Germany}
\author{Giulia Carini}
\affiliation{Fritz-Haber-Institut der Max-Planck-Gesellschaft, Faradayweg 4-6,14195 Berlin, Germany}
\author{Dmitry N. Chigrin}
\affiliation{DWI - Leibniz-Institut für Interaktive Materialien, Forckenbeckstr. 50, 52074 Aachen, Germany}
\affiliation{Institute of Physics (1A), RWTH Aachen University, 52074 Aachen, Germany}
\author{Alexander Paarmann}
\email{alexander.paarmann@fhi-berlin.mpg.de}
\affiliation{Fritz-Haber-Institut der Max-Planck-Gesellschaft, Faradayweg 4-6,14195 Berlin, Germany}

\date{\today}

\begin{abstract}
Phonon polariton modes in layered anisotropic heterostructures are a key building block for modern nanophotonic technologies. The light-matter interaction for evanescent excitation of such a multilayer system can be theoretically described by a transfer matrix formalism. This method allows to compute the imaginary part of the p-polarized reflection coefficient $\Im{r_{pp}}$, which is typically used to analyze the polariton dispersion of the multilayer structure, but lacks the possibility to access the layer-resolved polaritonic response. We present an approach to compute the layer-resolved polariton resonance intensity in aribtrarily anisotropic layered heterostructures, based on calculating the Poynting vector extracted from a transfer matrix formalism. Our approach is independent of the experimental excitation conditions, and fulfills an empirical conservation law. As a test ground, we study two state-of-the-art nanophotonic multilayer systems, covering strong coupling and tunable hyperbolic surface phonon polaritons in twisted \MoO~double layers. Providing a new level of insight into the polaritonic response, our method holds great potential for understanding, optimizing and predicting new forms of polariton heterostructures in the future.
%

\end{abstract}

\maketitle


\section{Introduction}

Layered heterostructures provide a versatile platform for the construction of nanophotonic devices, enabling extensive functionality of light propagating through nanoscale stratified systems \cite{Xia2014}. The tremendous progress reported using layered systems is significantly fueled by polaritons -- strong light-matter interaction featuring strongly localized, immense electric field strengths -- advancing a variety of nanophotonic fields such as optoelectronics \cite{He2013,Ross2014}, photovoltaics \cite{Fortin1982,Yu2013}, polaritonic optics \cite{Folland2018,Chaudhary2019,Passler2019a}, or sensing \cite{Rodrigo2015}. In particular, layered systems that are composed of strongly optically anisotropic polar crystals currently receive increasing interest due to their capability of supporting infrared polariton modes of high propagation directionality, so called hyperbolic phonon polaritons (hPhP) \cite{Jacob2014,Li2015,Dai2019,Passler2022,He2022}. While in isotropic polar crystals, phonon polaritons arise in the frequency region of negative permittivity between the transverse optical (TO) and longitudinal optical (LO) phonon modes, hPhPs in anisotropic crystals arise at frequencies where the permittivity is only negative along one (type I hyperbolic) or two (type II hyperbolic) principle crystal axes. Thin films of materials with out-of-plane anisotropy, such as hexagonal boron nitride (hBN), support volume-confined hPhPs, which have proven to enable subdiffraction imaging and hyperlensing \cite{Dai2015,Ferrari2015}. Materials with strong in-plane anisotropy, such as molybdenum trioxide (\MoO), on the other hand, support in-plane hyperbolic phonon polaritons (ihPhPs) featuring directional propagation in the surface plane. Only recently, the potential of these materials has captured attention, in particular demonstrated by the seminal work of several groups on twisted \MoO~layers \cite{Hu2020,Duan2020,Zheng2020,Chen2020,HerzigSheinfux2020}, where the twist angle enables control over the ihPhP wavefront geometries, propagation characteristics, and its topology. 

Advances in the field of polaritonic nanophotonics often are only feasible with the aid of a robust theoretical framework for the simulation of the optical response of the material system in question. For layered heterostructures, a $4 \times 4$ transfer matrix method (TMM) \cite{Passler2017a} has proven useful, providing the reflection and transmission coefficients as well as the local electric fields of a multilayer system consisting of any number of arbitrarily anisotropic materials. Furthermore, the analysis of the Poynting vector \SSS~allows for a layer-resolved calculation of the absorption and transmittance in the system even for fully anisotropic constituent materials \cite{Passler2020b}. However, polaritons typically are evanescent modes, that is, they feature in-plane momenta $k$ larger than the momentum of light in vacuum $k_0$, and thus cannot be accessed in a freespace excitation scheme. This condition for the excitation has to be accounted for in both the experimental as well as the theoretical observation of polaritons, and is, for instance, met in prism-coupling techniques such as the Otto geometry \cite{Otto1968,Passler2017,Folland2019} or the Kretschmann-Raether configuration \cite{Kretschmann1971}. While in particular the Otto geometry allows for a systematic, thorough study of phonon polaritons and has proven to be quite versatile \cite{Neuner2009,Passler2018,Ratchford2019,Passler2019a}, the intrinsic properties of the polariton modes in the sample are inevitably modified by the presence of the coupling prism. Other optical excitation techniques where large momenta are achieved by scattering off a nanoscale object, such as scattering-type scanning near field optical microscopy (s-SNOM) \cite{Huber2005,Novotny2006}, on the other hand, cannot fully be described theoretically using a $4 \times 4$ transfer matrix method, due to the deviation from a stratified system by the scattering source.

A common way to circumvent the specifics of the excitation method in the simulations is to calculate the optical response solely of the sample, with an excitation beam featuring artificially large in-plane momenta $k/k_0 > 1$. This evanescent wave excitation does not lead to physical results regarding the far-field reflectance, transmittance, or absorption, but has nevertheless proven insightful into the supported polariton mode dispersion \cite{Dai2014,Gubbin2019,Fali2019,Folland2018,AlvarezPerez2020,Passler2018}. In particular, the imaginary part of the p-polarized reflection coefficient $\Im{\rpp}$ peaks at frequencies where the system supports a polariton mode, thus providing a means to map out the instrinsic polariton dispersion. However, in layered heterostructures comprising several materials that support polaritons, the method of using $\Im{\rpp}$ only reveals the resonances of the overall system, while the relative distribution of the polariton resonance intensity across the layers remains inaccessible. For far-field excitations with $k/k_0 < 1$, a layer-resolved calculation framework for anisotropic multilayers has already been published \cite{Passler2020b}, but an equivalent method for evanescent excitation with $k/k_0 > 1$ has, to the best of our knowledge, not been discussed in literature so far.

Here, we present an empirical approach for the layer-resolved calculation of the relative intensity of polariton resonances in arbitrarily anisotropic layered heterostructures. The method of using $\Im{\rpp}$ for the determination of polariton dispersions, even though lacking a thorough theoretical justification so far, has been successfully and continuously used for several years. We build on this empirical knowledge, expanding the established method by a layer-resolved calculation based on the Poynting vector obtained from a $4 \times 4$ TMM that is implemented in an open-access computer program \cite{Passler2022a}. We demonstrate our method by calculating the layer-resolved polariton resonances in two state-of-the-art polaritonic systems, covering strongly coupled SPhPs in an aluminum nitride (AlN) / silicon carbide (SiC) heterostructure, and tunable ihPhPs in twisted \MoO~layers on a quartz (\SiO) substrate. Fulfilling an empirical conservation law, our method provides insight into the relative intensity of the polariton resonances in the different layers of the sample system.

\section{Method}

The TMM we employ in this work has been described in detail previously \cite{Passler2017a}. For the calculation of the layer-resolved polaritonic response of the sample system, we further use an extended formalism based on the TMM \cite{Passler2020b}, providing the time-averaged Poynting vector $\vec{\SSS}^p_i(z)$ for p-polarized incident light, in layer $i$, at position $z$:
\begin{align}
\vec{\SSS}^p_i(z) = \frac{1}{2} \text{Re} \left[ \vec{\E}^p_i(z) \times \vec{\HH}_i^{p*}(z) \right],
\label{eq:Svec}
\end{align}

where $\vec{\E}_i(z)$ and $\vec{\HH}_i(z)$ are given elsewhere \cite{Passler2020b}. Note that this formalism is originally designed for propagating incident light with $k/k_0 < 1$. Further on (Eq. 5), we extend the method to evanescent excitation with $k/k_0 > 1$. The coordinate system is chosen such that the $z$-axis points along the surface normal, the exciting light beam is incident in the $x$-$z$-plane, and the origin of the coordinate system lies in the interface plane between the semi-infinite incident medium ($i=0$) and the first layer ($i=1$). The multilayer system comprises $N$ layers of thicknesses $d_i$, and layer $i=N+1$ is the semi-infinite substrate. Because polaritons are only excitable by p-polarized light \cite{Passler2019a}, we omit the specification of the incoming polarization in the following, referring always to p-polarization.

In order to calculate the transmittance up to layer $i$ and position $z$, the $z$-component of the Poynting vector at the corresponding position is normalized by the $z$-component of the Poynting vector of the incoming excitation beam $\SSS_{\text{inc},z}$:
\begin{align}
\T_i^{a}(z) &= \frac{\SSS_{i,z}(z)}{\SSS_{\text{inc},z}},
\label{eq:Ti}
\end{align}

and the transmittance $\T$ into the substrate $i=N+1$ at the interface with layer $N$ is given by:
\begin{align}
\T = \frac{\SSS_{N+1,z}(D)}{\SSS_{\text{inc},z}},
\end{align}

where $D=\sum_{i=1}^{N} d_i$ is the thickness of the multilayer system. Using Eq. \ref{eq:Ti}, the layer-resolved absorption can be calculated as follows:
\begin{align}
\begin{split}
\A_i &= \T_i(d_{1..i-1}) - \T_i(d_{1..i-1} + d_i),
\end{split}
\label{eq:Ai}
\end{align}
where $d_{1..i-1} = \sum_{i=1}^{i-1} d_i$ is the thickness of all layers through which the incident light has propagated before reaching the layer $i$.

For a propagating excitation beam with $k/k_0 < 1$, $\SSS_{\text{inc},z}$ is real-valued, as specified in Eq. 22 of reference \cite{Passler2020b}, and $\A$ and $\T$ correctly describe the absorption and transmission, respectively. For an evanescent incident beam with $k/k_0 > 1$, however, $\SSS_{\text{inc},z}$ is purely imaginary, since an evanescent beam features no net energy flow in $z$-direction. As a consequence, for evanescent excitation, this would lead to a zero denominator in Eq. \ref{eq:Ti}. Here, we therefore normalize to the imaginary part of $\SSS_{\text{inc},z}$ instead, as can be calculated from Eq. 22 of reference \cite{Passler2020b} with the following modification:
\begin{align}
\vec{\SSS}_{\text{inc}} &= \frac{1}{2} \text{Im} \left[ \vec{\E}_{\Rightarrow,0}(0) \times \left( \kvec_{01} \times \vec{\E}_{\Rightarrow,0}(0)\right)^* \right],
\label{eq:Sinc}
\end{align}
where $\Rightarrow$ denotes the forward propagating (incident) mode, and $\vec{k}_{01}$ is the wavevector of the p-polarized incident beam.

\begin{figure*}[!ht]
\includegraphics[width = \textwidth]{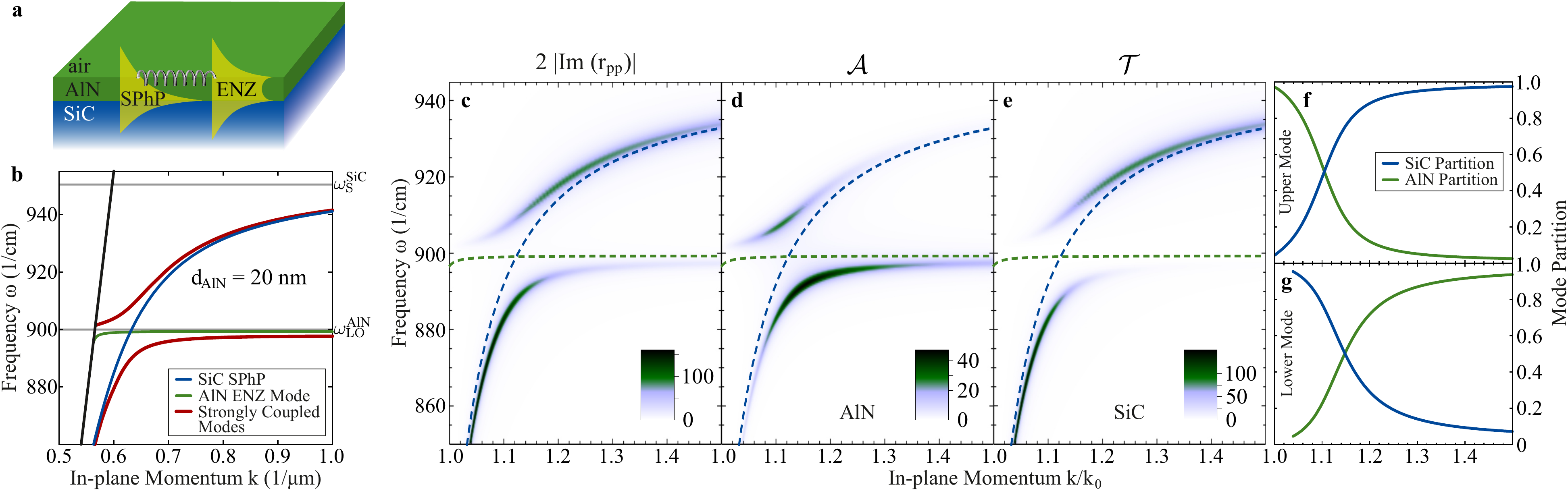}
\captionof{figure}{\textbf{Strong coupling between an AlN ENZ mode and a SiC SPhP.} 
	\textbf{a} Sketch of the AlN/SiC structure, illustrating the strong coupling of a SPhP of a bare SiC substrate and an ENZ mode of a freestanding AlN film. \textbf{b} Analytical dispersion of the uncoupled SiC SPhP (blue line) and AlN ENZ mode (green line), as well as the resulting strongly coupled modes in the heterostructure (red lines) featuring an avoided crossing. \textbf{c} Dispersion of the strongly coupled modes obtained by calculating the total resonance intensity $\Im{r_{pp}}$. \textbf{d,e} Layer-resolved distribution of the resonance intensity in AlN and SiC, respectively. In c,d, and e, the analytical dispersions of the uncoupled SiC SPhP (blue lines) and AlN ENZ modes (green lines) are plotted for reference. \textbf{f,g} Mode partition of the AlN film and the SiC substrate for the upper and the lower dispersion branch, respectively.}
\label{fig1}
\end{figure*} 

We note that this modified normalization to the imaginary part is, similar to the use of $\Im{\rpp}$, empirically motivated. Nonetheless, the layer-resolved "absorption" calculated according to Eq. \ref{eq:Ai} conveniently reflects the relative intensities of a polariton mode present in the different layers of a multilayer structure, as we will demonstrate in the following section. Furthermore, please note that, analogously to $\Im{\rpp}$, $\T$ and $\A$ take values larger than 1 in the case of $k/k_0 > 1$, rendering the use of the terms "transmittance" and "absorption" inadequate. Therefore, in the following we refer to the quantities simply by their mathematical symbols.

Strikingly, the sum of the layer-resolved quantities $\A_i$ and $\T$ fulfill, as we have numerically verified for a broad variety of test cases, the following conservation law:
\begin{align}
2 \Im{r_{pp}} = \sum_{i=1}^{N} \A_i  + \T,
\end{align}

where we calculate $r_{pp}$ employing a TMM \cite{Passler2017a}. This equation constitutes the conservation between the resonance intensity distributed between the layers of the system described by $A_i$ and $\T$, and the overall resonance intensity, here found to be $2 \Im{r_{pp}}$. 

In the following, we will apply our method to two sample systems that have been discussed in literature before, demonstrating that our results are not only in accordance with previous findings, but also provide additional insight into the resonance behavior of polariton modes in layered heterostructures. 

\section{Strongly Coupled ENZ Polaritons}
At frequencies close to zero crossings of the real part of the dielectric permittivity \epsi, a material features epsilon-near-zero (ENZ) light propagation with remarkable properties of the ENZ photonic modes, such as high emission directionality \cite{Enoch2002,Kim2016}, enhanced nonlinear-optical conversion efficiency \cite{Argyropoulos2012,Suchowski2013}, and tunneling through narrow distorted waveguide channels \cite{Silveirinha2007,Edwards2009}. In a polar crystal, ENZ conditions are met at the LO phonon frequency \wLO, and an ENZ polariton can be found in subwavelength-thin polar crystal films \cite{Vassant2012,Nordin2017,Campione2015}. However, a thin-film ENZ polariton is a non-propagating mode due to its intrinsically flat dispersion close to \wLO, thus hindering its usability for effective nanoscale communication applications. This limitation can be overcome by strongly coupling an ENZ polariton to a propagating SPhP, as has been demonstrated for an aluminum nitride (AlN) thin film / silicon carbide (SiC) heterostructure \cite{Passler2018}, see Fig.~1a. By combining the advantages of the constituent uncoupled modes, the resulting ENZ-SPhPs feature strong electrical field enhancement characteristic for ENZ modes, while maintaining a propagative character typical for SPhPs. 

The dispersions of both the uncoupled AlN ENZ mode (green line) and the SiC SPhP (blue line) as well as the strongly coupled modes (red lines) are plotted in Fig.~1b, calculated with an analytical formula for a three-layer system \cite{Burke1986,Campione2015}. Characteristically for strong coupling, the ENZ-SPhP dispersion lines exhibit an avoided crossing, while approaching the dispersion lines of the uncoupled modes with increasing distance to the dispersion crossing point. Accordingly, the mode nature along each of the strongly coupled mode dispersions undergoes a transition across the avoided crossing, while at the avoided crossing, both strongly coupled modes have identical characteristics such as electric field enhancement and spatial confinement \cite{Passler2018}, sharing equal measures of both uncoupled modes. In order to verify and visualize this transition of mode nature across the strong coupling region, we here apply our method to calculate the polariton resonance intensity in the AlN/SiC heterostructure resolved for each layer.

The overall polaritonic response of the material system can be obtained by calculating $\Im{r_{pp}}$, as it is shown in Fig.~1c, where the entire dispersions of both strongly coupled modes are reproduced. The layer-resolved calculations obtained from our method are plotted in Fig.~1d ($\A$ in AlN) and Fig.~1e ($\T$ in SiC). For both layers, only parts of the same dispersion lines as for $\Im{r_{pp}}$ are obtained. In the AlN film (Fig.~1d), the resonance intensity is strongest in close proximity to the AlN ENZ mode (green line), whereas the intensity fades out along the SiC SPhP (blue line). In the SiC substrate (Fig.~1e), on the contrary, the resonance intensity is most pronounced along the SiC SPhP and almost no intensity can be found along the AlN ENZ mode. This relative intensity distribution between the different layers reflects the respective partial mode nature along the dispersion, changing from the AlN ENZ mode to the SiC SPhP and vice versa. This behavior can be demonstrated by quantifying the mode partition \PP~as follows:
\begin{align}
\PP_i = \frac{\A_i}{2 \Im{r_{pp}}},
\end{align}
and evaluating $\PP_{\text{SiC}}$ and $\PP_{\text{AlN}}$ (blue and green lines) along both dispersion branches of the strongly coupled polariton modes, as shown in Fig.~1f and g, respectively. Clearly, along both branches the mode nature undergoes the aforementioned transition, with a crossing point where the mode exhibits AlN ENZ and SiC SPhP features in equal measures. Notably, this crossing point sits at slightly different in-plane momenta for the upper and the lower branch, corresponding to the momentum where the uncoupled mode dispersions are equidistant to the respective branch in frequency-momentum space. 

An alternative approach to obtain the relative mode distribution in the multilayer system would be to calculate the layer-resolved absorption for excitation with a propagating wave ($k/k_0 < 1$) via Otto-type prism coupling. However, in this scheme, the relative absorption of the polariton modes is distorted by the coupling prism, because the AlN ENZ and the SiC SPhP modes feature distinct critical gaps of optimal coupling conditions. In contrast, our approach is free of the influence of the excitation method, revealing consistent additional information about the mode nature of the strongly coupled modes in the AlN/SiC heterostructure.  

\begin{figure*}[!ht]
\includegraphics[width = \textwidth]{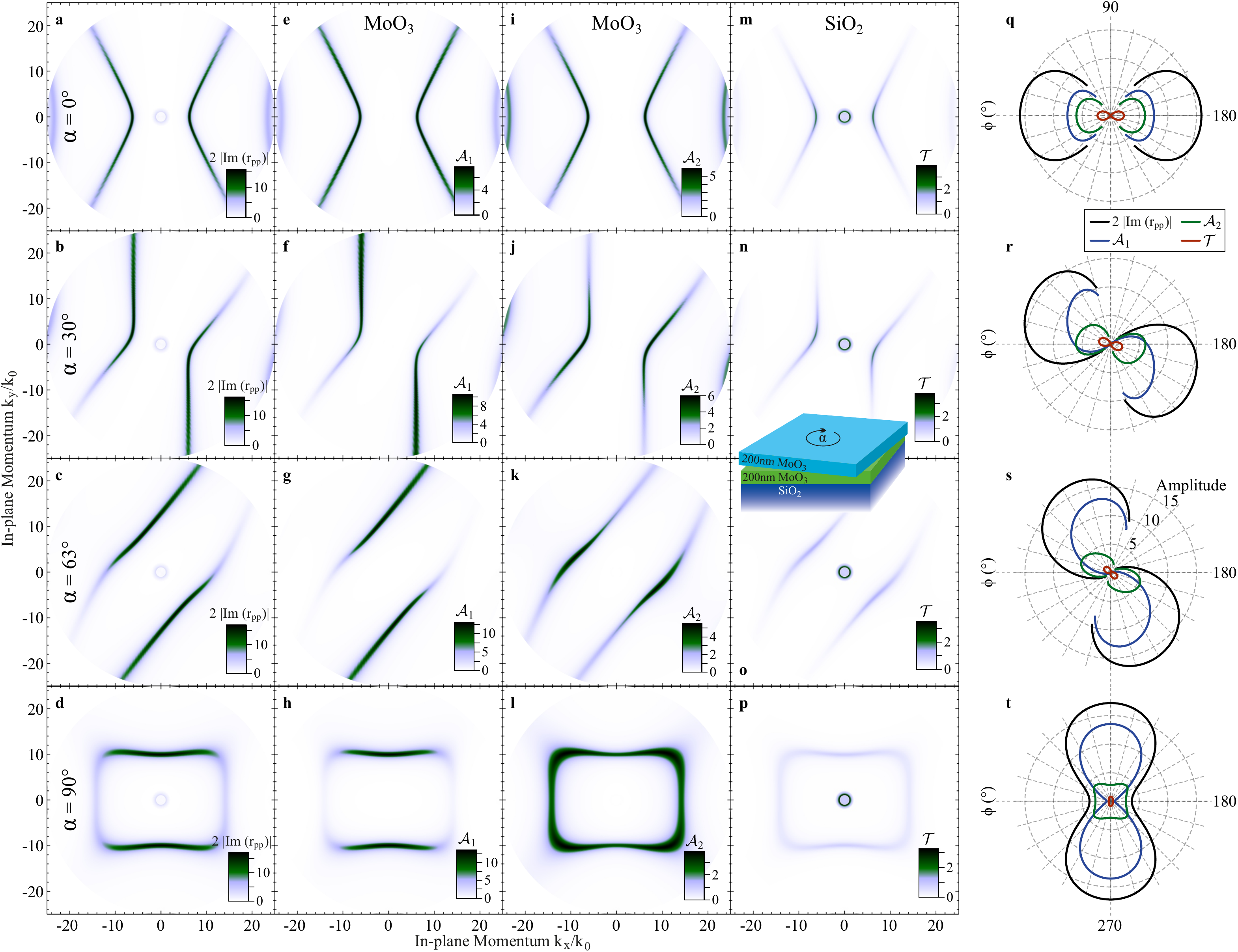}
\captionof{figure}{\textbf{Tunable phonon polaritons in twisted \MoO~layers.} 
	\textbf{a-d} $\Im{r_{pp}}$ as a function of in-plane momenta $k_x/k_0$ and $k_y/k_0$ for a \nmetr{200} \MoO/\nmetr{200} \MoO/\SiO~heterostructure, as illustrated in the inset, at four different twist angles $\alpha=0, 30, 63, \dg{90}$ of the upper \MoO~layer, respectively. The calculations reveal a topological transition at the magic twist angle $\alpha^*=\dg{63}$ from an ihPhP to an elliptical SPhP. \textbf{e-h} Layer-resolved resonance intensity $\A_1$ in the upper and \textbf{i-l} in the lower \MoO~layer, \textbf{m-p} $\T$ in the \SiO~substrate, and \textbf{q-t} polar plots of the resonance intensities of all four quantities along the dispersion of the first-order SPhP mode, each at four different twist angles $\alpha$, respectively.}
\label{fig1}
\end{figure*}

\section{In-Plane Hyperbolic Polaritons in Twisted \MoO~Layers}
In-plane hyperbolic phonon polaritons (ihPhPs) are supported on polar crystals with in-plane hyperbolicity, that is, at frequencies where $\Re{\epsi_x} \Re{\epsi_y} < 0$ (with the crystal surface lying in the $x$-$y$-plane). The dispersion of ihPhPs take the form of a hyperbola in the surface plane, oriented such that the hyperbola minimum lies on the crystal axis along which $\Re{\epsi} < 0$, whereas no solution is supported along the perpendicular surface direction where $\Re{\epsi} > 0$. Therefore, ihPhPs intrinsically feature a strong propagation directionality.  At frequencies where both in-plane permittivity tensor elements are negative, on the other hand, the dispersion describes an ellipse, and the resulting SPhP can propagate along any direction in the surface plane.

Recently, it has been demonstrated that by stacking and twisting two \MoO~layers, the propagation direction of the supported surface polaritons becomes configurable as a function of the twist angle $\alpha$ \cite{Chen2020,Hu2020,Duan2020}. Furthermore, at a specific, frequency-dependent magic angle, the surface polariton performs a topological transition from a hyperbolic to an elliptical dispersion. The overall change in propagation direction and topology as a function of $\alpha$ is well-captured by $\Im{r_{pp}}$, as is reproduced in Fig.~2a-d in perfect agreement with literature. At twist angles $\alpha = \dg{0}$ and \dg{30} (Fig. 2a,b respectively), the polariton is hyperbolic, and the propagation direction rotates with $\alpha$. At the magic angle $\alpha^*=\dg{63}$, the dispersion transitions from hyperbolic to elliptical, resulting in flattened dispersion lines that exhibit diffractionless and low-loss directional polariton canalization \cite{Hu2020}. Finally, at $\alpha=\dg{90}$ (Fig. 2d), the topological transition is completed and the stacked system features an "elliptical" dispersion (that is, finite in all in-plane directions) of almost rectangular shape. 

In order to reveal the optical response resolved for each material layer of the twisted heterostructure, we employ our formalism to calculate $\A_1$ and $\A_2$ for the two \MoO~layers, and $\T$ for the \SiO~substrate (the system is sketched in the inset in Fig.~2o). The resonance intensities $\A_1$ and $\A_2$ for the four twist angles $\alpha=0,30,63,\dg{90}$ in the first and second \MoO~layers are shown in Fig.~2e-h and 2i-l, respectively, and the resonance intensity $\T$ in the substrate is plotted in Fig.~2m-p. Finally, the resonance intensity peak value along the dispersion of the first-order mode is shown in polar plots in Fig.~2q-t. Note that the curves are not continuous for $\alpha=0,30,$ and \dg{63} because of the finite plot range and the divergent nature of the dispersion. 

Clearly, the maximum resonance intensity is strongest in the first \MoO~layer and decreases towards the substrate (Fig.~2q-t). As a consequence, rotating the first layer dominates the overall maximum intensity along the dispersion in $\Im{r_{pp}}$ (black lines), which rotates with $\alpha$. The same is true for \T~in the isotropic \SiO~substrate (red lines). The intensity maxima of $\A_1$ and $\A_2$ in the first and second \MoO~layer, however, follow the orientation of the optical axis in the respective layer, where in the first layer (blue lines), the maximum is shifted clockwise in the direction of the twist rotation, while in the second layer (green lines), the maximum is only mildly rotated. This leads to strongly asymmetric intensity distributions along the dispersion in both \MoO~layers for the hyperbolic region, that is, at twist angles $\alpha=30$ and \dg{63} (Fig.~2f,j and g,k, respectively). At $\alpha=\dg{90}$, finally, the intensity maximum is oriented along the y-axis and arises mostly from the first \MoO~layer, while the small fraction of resonance intensity along the x-axis solely originates in the second layer.

By resolving the spatial origin of the resonance intensity layer by layer, our method reveals that the partial resonance intensity in each \MoO~film is oriented along the respective polariton-active crystal axis. However, due to the presence of the respective other \MoO~layer, the partial response in each \MoO~film can feature strongly asymmetric azimuthal intensity distributions, depending on the twist angle $\alpha$. Thus, the polariton modes of the individual films are modified by the presence of the adjacent twisted \MoO~film, while not featuring full hybridization, as has been observed in the previous example system. Finally, the resulting polariton mode in the full system can be seen as the sum of these partial polaritonic responses in each \MoO~layer. Revealing this layer-resolved information, our method therefore provides a deeper analysis of the supported ihPhP modes for each topological state in the twisted \MoO~double layer heterostructure, and may even accomplish the guiding principles for engineering the dispersion.

\section{Discussion}
We have introduced here an empirical approach to analyze the layer-resolved intensities of evanescent modes in heterostructures. Yet, it remains unresolved how to embed such a method into a solid theoretical framework, where for instance energy conservation is rigorously traceable and absorption and transmission take physical values $<1$.  The mode partition in the air/AlN/SiC strong coupling system, Fig.~1f, may give a hint, though, on how this could be achieved. Consider that with evanescent wave excitation also the reflected wave is evanescent and cannot transport any energy, that is, the reflectance is $0$ by definition. Then, the SiC partition would actually define the transmission while the AlN partition defines the absorption. In such a picture, reflectance, absorption and transmission take physical values, i.e. $\R=0$, $\A, \T \leq 1$ and $\R + \A + \T = 1$. This would still hold true also in multilayer systems, with more than one layer contributing to the total absorption. While this analogy is intriguing, it is beyond the scope of this work to rigorously connect these considerations to the well-established physics of propagating plane waves.

Nonetheless, our empirical method reveals unprecedented details on the polariton distribution in multilayer systems at low computational cost. Following the recent success of twisted double layer structures, we anticipate high demand for modeling forthcoming twisted multilayer concepts. Here, our approach could provide comprehensive data that may significantly help to identify the guiding principles for designated design goals. If additionally the relevant physics is driven by the polariton intensity in a specific layer or at a given interface of the structure, as for example expected for polariton-driven chemistry, the relevance of our layer-resolved analysis is enhanced even further. As a natural extension, it would be highly desirable to be able to quantitatively connect the empirical results obtained here to experimentally accessible quantities, as for instance the scattering amplitude and phase in nano-FTIR or s-SNOM, which would enable much enhanced data analysis capabilities for multilayer structures.

\section{Conclusion}
In this work, we have presented an empirical approach for the layer-resolved analysis of the resonance intensity of polariton modes in arbitrarily anisotropic, birefringent, and absorbing multilayer media. Our method builds on the empirical approach of calculating the imaginary part of the reflection coefficient $\Im{r_{pp}}$ for evanescent wave excitation that has been successfully used in literature for several years. The resulting layer-resolved resonance intensities that we calculate from the Poynting vectors obtained from a TMM \cite{Passler2017a,Passler2020b} fulfill an empirical conservation law, balancing the resonance intensity expressed in $\Im{r_{pp}}$ with the sum of the resonance intensities in each system layer. The presented method is implemented in an open-access computer program \cite{Passler2022a}. 

As case studies, we applied our approach to the analysis of two recently studied nanophotonic systems featuring strong coupling between an ENZ and a propagating SPhP mode and the modulation of the propagation direction and the topological state of ihPhPs, revealing yet undiscovered details about the supported polariton modes. By allowing to analyze any multilayer system independent of the excitation scheme, our method holds great potential for understanding, optimizing and predicting new forms of polariton heterostructures in the future.

\section{Acknowledgments} 
We thank M. Wolf, S. Wasserroth and R. Ernstorfer (FHI Berlin) for careful reading of the manuscript and M. Wolf and the Max Planck Society for supporting this work.

\bibliographystyle{apsrev4-2}
\bibliography{twisted_MoO3_layers} 

\end{document}